# Digital Discovery

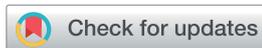

# Semi-supervised machine learning workflow for analysis of nanowire morphologies from transmission electron microscopy images†

Shizhao Lu, [ID] [a] Brian Montz, [b] Todd Emrick [ID] [b] and Arthi Jayaraman [ID] *[ac]

In the field of materials science, microscopy is the first and often only accessible method for structural characterization. There is a growing interest in the development of machine learning methods that can automate the analysis and interpretation of microscopy images. Typically training of machine learning models requires large numbers of images with associated structural labels, however, manual labeling of images requires domain knowledge and is prone to human error and subjectivity. To overcome these limitations, we present a semi-supervised transfer learning approach that uses a small number of labeled microscopy images for training and performs as effectively as methods trained on significantly larger image datasets. Specifically, we train an image encoder with unlabeled images using self-supervised learning methods and use that encoder for transfer learning of different downstream image tasks (classification and segmentation) with a minimal number of labeled images for training. We test the transfer learning ability of two self-supervised learning methods: SimCLR and Barlow-Twins on transmission electron microscopy (TEM) images. We demonstrate in detail how this machine learning workflow applied to TEM images of protein nanowires enables automated classification of nanowire morphologies (*e.g.*, single nanowires, nanowire bundles, phase separated) as well as segmentation tasks that can serve as groundwork for quantification of nanowire domain sizes and shape analysis. We also extend the application of the machine learning workflow to classification of nanoparticle morphologies and identification of different type of viruses from TEM images.

## Introduction

Researchers working with nanomaterials find that the materials' structural characterization is a key step in the discovery of novel functional materials.[1,2] For most researchers, microscopy imaging is the first and, in many cases, only accessible means to obtain structural information. Depending on the length scale of interest, commonly used techniques include optical microscopy, transmission electron microscopy (TEM), scanning electron microscopy (SEM), and atomic force microscopy (AFM). Regardless of the technique, each microscopy image is associated with the chemical composition of the material, region/section of the material that is imaged, and the processing conditions for the imaging. The resulting images may resemble a heat map of intensity highlighting object(s) owing to the selective staining of some species/sections in the sample.[3] Analyzing these microscopy images requires domain knowledge to interpret (*e.g.*, classify images with morphology labels) and/or to detect nuances in different images (*e.g.*, identification of defects, changes in intensity, *etc.*) as the composition or the focus of the image or the processing condition is changed. Thus, manual interpretation is often time-consuming, labor-intensive, and prone to human subjectivity in interpretation. Therefore, machine learning (ML) has become a valuable tool that can replace time-consuming and subjective manual interpretation of microscopy images with automated, objective, and fast analysis.[4–8]

Recent advances in deep learning have led to a surge of applications in electron microscopy image analysis for a diverse set of tasks in two main categories: discriminative and generative. Discriminative tasks are tasks like morphology/phase classification,[9–12] particle/defect detection,[13–16] image quality assessment,[17–19] and segmentation[20–25] where the objective is quantified by how well the model can distinguish (1) between images or (2) between objects and their background. Generative tasks include structure reconstruction,[26–28] super resolution,[29–31] autofocus[32] and denoising[33,34] where the objective is generation of images with certain desired traits.

[a]Department of Chemical and Biomolecular Engineering, University of Delaware, 150 Academy St., Newark 19716, Delaware, USA. E-mail: arthij@udel.edu

[b]Department of Polymer Science and Engineering, University of Massachusetts Amherst, Amherst 01003, Massachusetts, USA

[c]Department of Materials Science and Engineering, University of Delaware, 201 DuPont Hall, Newark 19716, Delaware, USA

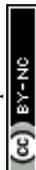

Development of modern ML models benefits in performance from the procurement of big datasets related to a specific task. To bypass the need to collect large training data and reduce the time needed to train the ML model from scratch on that large data, researchers use 'transfer learning' techniques.[35] Transfer learning involves leveraging the knowledge of a model previously trained using large training datasets to create a new model for another related task. For example, a model that has been trained on ImageNet,[36] a large dataset of 1.2 million photographic images of macroscopic objects, can be transferred to learn how to analyze images in another more specific domain [e.g., medical image analysis[37,38] and electron microscopy image analysis in material science[4-8]]. The success of transfer learning in the field of image analysis has paved the way for accessibility to pretrained image learning models for the general public without requiring large computational resources or big data to train from scratch.[39]

Transfer learning for microscopy image analysis tasks has traditionally relied on convolutional neural network (CNN) models[40] which convert input images into feature maps that hold information for image classification (e.g., assigning a morphology label to a microscopy image) and detection of objects in the image (e.g., identification of a nanoparticle aggregate or domain). In transfer learning, the weights of some layers of the pre-trained CNN are kept as constants and only the weights for the outermost layer are retrained with images for the specific task at hand. In most implementations of transfer learning, the microscopy image dataset for the specific task has to be labeled (i.e., supervised learning) before training the outermost layer. However, CNN models trained on one type of supervised tasks (classification, segmentation, or object detection) can typically only be transferred to the same supervised task for another image dataset which limits the generalized applicability of transfer learning. Further, the typical size of image datasets needed for training ranges from thousands to hundreds of thousands of images, even with transferred models, and the labeling of these large set of images is challenging and prone to error, with a recent study noting (on average) 3.3% labeling errors in large open source datasets.[41]

To overcome limitations of labeling, semi-supervised training workflow is another option which typically consists of an unsupervised training of feature encoder requiring no manual labeling and a supervised training of specific downstream task model requiring manual labels.[42,43] Chowdhury et al. developed a semi-supervised approach consisting of a feature extractor, a feature selector, and a classifier to classify different dendritic microstructures.[9] Peikari et al. developed a cluster-then-label semi-supervised approach for classifying pathology images.[44] A school of generative architectures called autoencoders have also been used in obtaining pretrained feature maps of images.[45,46] An autoencoder architecture involves training of an encoder to condense the information from the original image to a low-dimensional feature map, and a decoder that tries to reconstruct the original image from the feature map. More recently, self-supervised learning of images has emerged as a new form of label-free, unsupervised training.[47] In a recent review, Liu et al. attribute the more competitive status of self-supervised learning compared to autoencoders in classification scenarios to the more closely aligned learning goal of self-supervised learning modules to that of vision tasks targeting high-level abstraction such as classification and object identification.[48] Through self-supervised training, the ML model learns a representation of an image by maximizing the similarity between two different transformed versions of the same image. While supervised model training is assisted by the labels associated with each image, self-supervised model training does not rely on labels and learns from the underlying features of the images. The performance of self-supervised transfer learning has been shown to be comparable with supervised transfer learning in big data medical image classification.[49,50] For example, Azizi et al. have achieved 70% classification accuracy on dermatology images (using ~450 000 images for training) with self-supervised transfer learning[49] and Ciga et al. have achieved 78% classification accuracy on a diverse set of histopathology images (using ~40 000 images for training) with self-supervised transfer learning.[50] In contrast to the medical imaging field which traditionally has large data sets, researchers in the soft materials domain handle much smaller datasets and have a more diverse range of image analysis tasks. Thus, for self-supervised transfer learning to be accessible to researchers in soft materials, it has to be adapted to small dataset sizes and be able to handle multiple tasks (e.g., classification and segmentation).

Transfer learning from CNNs trained with supervised methods has been utilized in nanomaterial classification task in recent years. Moderres et al. have used transfer learning to classify SEM images belonging to different nanomaterial subcategories like particles, patterned surfaces, nanowires, etc.[10] Having an unbalanced dataset, they observed higher accuracies for categories that have fewer images. They made the comment that some categories that have fewer images performed better because the features in those categories were distinct and sufficiently clear to be learned by the network. While some categories with higher number of images suffered from having indistinct features. Their dataset size was ~18 000 in total with smallest category containing ~150 images, and largest category containing ~4000 images. Luo et al. used transfer learning to classify carbon nanotube or fiber morphologies on an image dataset containing 600 images per morphology class.[11] They were able to achieve 91% average accuracy on a four-class dataset, and 85% average accuracy on an eight-class dataset. Matuszewski and Sintorn recently compared the accuracy of different CNN architectures (with transferred weights or trained from scratch) on identifying various viruses from TEM images.[51]

In this article, we present an automated, label-efficient transfer learning workflow incorporating self-supervised pre-training that aims to classify nanomaterial morphologies in microscopy images with high accuracy after training on only a handful of carefully labeled microscopy images. We focus on a semi-supervised transfer learning workflow to perform automated classification and segmentation of TEM images taken from one class of nano materials – protein and peptide

nanowires – which are used in a wide variety of applications including flexible electronics,[52] energy harvest,[53] and chemical sensing.[54] In all these applications, the nanowire morphologies (*e.g.*, dispersed, aggregated, percolated, phase separated) dictate their performance. In this work, we use TEM images from assembled, synthetically engineered, peptide nanowires and biologically derived (from *Geobacter sulfurreducens*) protein nanowires[55] to demonstrate a semi-supervised transfer learning workflow that shows high accuracy in classification and segmentation of these images with <1000 generic unlabeled training images and <10 task-specific labeled images per morphology class. We also demonstrate the broader applicability of our machine learning workflow by applying it to two additional TEM image analysis tasks – for classification of nanoparticle morphologies and for identification of virus types from their TEM images.

## Results and discussion

### Semi-supervised machine learning workflow

We illustrate the conceptual workflow of semi-supervised transfer learning for microscopy images in Fig. 1. First, a generic image learning model, an encoder, undergoes self-supervised training (*i.e.*, no labels required during training) on a dataset of generic microscopy images called CEM500k,[56] an open-access electron microscopy image dataset curated from various imaging modalities characterizing cellular or biomaterial structures by Conrad and Narayan. We transfer the trained encoder to transform images into feature maps (*i.e.*, distilled and encoded representations of images) for training of downstream tasks. We demonstrate the semi-supervised machine learning workflow with a detailed example of transfer learning of nanowire morphologies from generic TEM images. We start by training the encoder with self-supervised methods on generic TEM images (Fig. 1A). We implement two self-supervised training methods: SimCLR[47] and Barlow-Twins.[57] Both methods start by taking a batch of images and generating two randomly augmented images for each image by performing random color/hue/contrast changes, and randomly crop a portion of the image. The augmented images are then turned into feature maps by an encoder with ResNet50 (ref. 58) architecture. The feature maps are input into a projector with three layers of fully connected neurons to generate projections of each image. The projections are then used to calculate and minimize the loss function to train both the encoder and the projector. Through maximizing the similarity between two augmented images of the same image, the encoder is trained to produce feature maps that can represent the images more accurately. The difference between the two methods – SimCLR[47] and Barlow-Twins[57]- is in the loss function. The loss function of SimCLR method aims to maximize the calculated cosine similarity of projections from the "true" pairs of augmented images from the same image and minimize that of the "false" pairs of augmented images from different images. The loss function of SimCLR method has dependence on the batch size and contrast between images; larger batch size and higher contrast theoretically gives higher ability of discerning "true" from "false" pairs.

The loss function of Barlow-Twins method aims to minimize the redundancy in the representation of the projection by tuning the cross-correlation matrix of projections from the same image to be an identity matrix. The equations of the two loss functions are presented in the methods section with a more detailed explanation. The trained encoder is then transferred to learn the protein/peptide nanowire morphologies (Fig. 1B). For the classification task, we use the simplest linear classifier consisting of four neurons equivalent to the number of morphology classes to classify the feature maps. For the image segmentation task, we use U-Net,[59] an established deep learning architecture that has been shown to outperform traditional segmentation methods.[23,60,61] The original U-Net architecture consists of an encoder and a decoder trained to create accurate segmentation of input images. Instead of the original U-Net architecture, we use our trained encoder with trained weights to establish skip connections between our encoder and a decoder with random initialized weights.

### Peptide/protein nanowires – morphology classification

Protein/peptide nanowires exhibit one of four morphologies when dispersed in solvent – singular (*i.e.*, isolated nanowire), dispersed (*i.e.*, isolated collection of multiple nanowires), network (*i.e.*, percolated nanowires), and bundle morphologies. Materials with dispersed nanowires are desired for mechanical reinforcement,[62] while materials with network morphologies are desired for improving conductivity.[52] The singular, dispersed, and network morphologies in this work arise from assembly of synthetic oligopeptides shown in Fig. 2A; the bundle morphologies represent aggregates of protein nanowires harvested from *Geobacter sulfurreducens*. 100 images from each morphology are employed (Fig. 2B). The magnification of the morphology images varies from image to image, but the length scales are on the same order of magnitude as indicated from the scale bars in Fig. 1B. Because the interest of our study is the type of morphology rather than the length scale of the morphology, we do not include the scale bars in the images for training the machine learning models. Due to differences in the peptide/protein nanowire chemistry, solvent condition and magnification, the object-background contrast in each morphology image is different. As the dispersed and network morphologies are harder to visually distinguish, we manually label the nanowires in the images through Microscopy Image Browser (MIB).[63] This manual labeling serves two purposes: (1) to provide pixel-level quantification of percolation (in network) or lack thereof (in dispersed) and (2) to provide ground truth labels of nanowires for the segmentation task. A percolation analysis of the clusters of manually labeled nanowires distinguishes the networks (percolated) and dispersed (not percolated) nanowire morphologies. Except for the state-of-the-art (SotA) encoder which we obtained as a published open-access encoder trained with SimCLR method on the full ImageNet dataset (1.26 million images), other models are trained with optimized hyperparameters detailed in the method section. To show whether transferring learned weights from a domain-specific dataset gives better model performance, we trained the two self-supervised encoders on 832 generic TEM images or 832 generic everyday photographic images from ImageNet, both at resolution $224 \times 224$. We report the classification accuracies from the linear classifier on the feature maps of the test sets (Fig. 2C). The classification accuracy is defined as a ratio of the number of correct morphology class predictions *e.g.*, an image of dispersed morphology predicted as dispersed morphology, over the total number of the test cases. When trained with generic TEM images, the Barlow-Twins method outperforms SimCLR method. When trained with Barlow-Twins method, transferring from domain-specific images, *i.e.*, TEM images, brings higher performance than transferring from everyday images. However, when trained with SimCLR method, transferring from domain-specific images underperforms transferring from images of other domains. We believe that SimCLR performs worse when trained on generic TEM images due to the reduced contrast in generic TEM images compared to that in ImageNet images. Strikingly, feature maps obtained from the Barlow-Twins-TEM encoder obtain >90% classification accuracy when trained with just 8 labeled images per class. With more numbers of labeled images, feature maps obtained from the Barlow-Twins-TEM encoder achieve comparable classification accuracy to that of feature maps obtained from the SotA encoder.

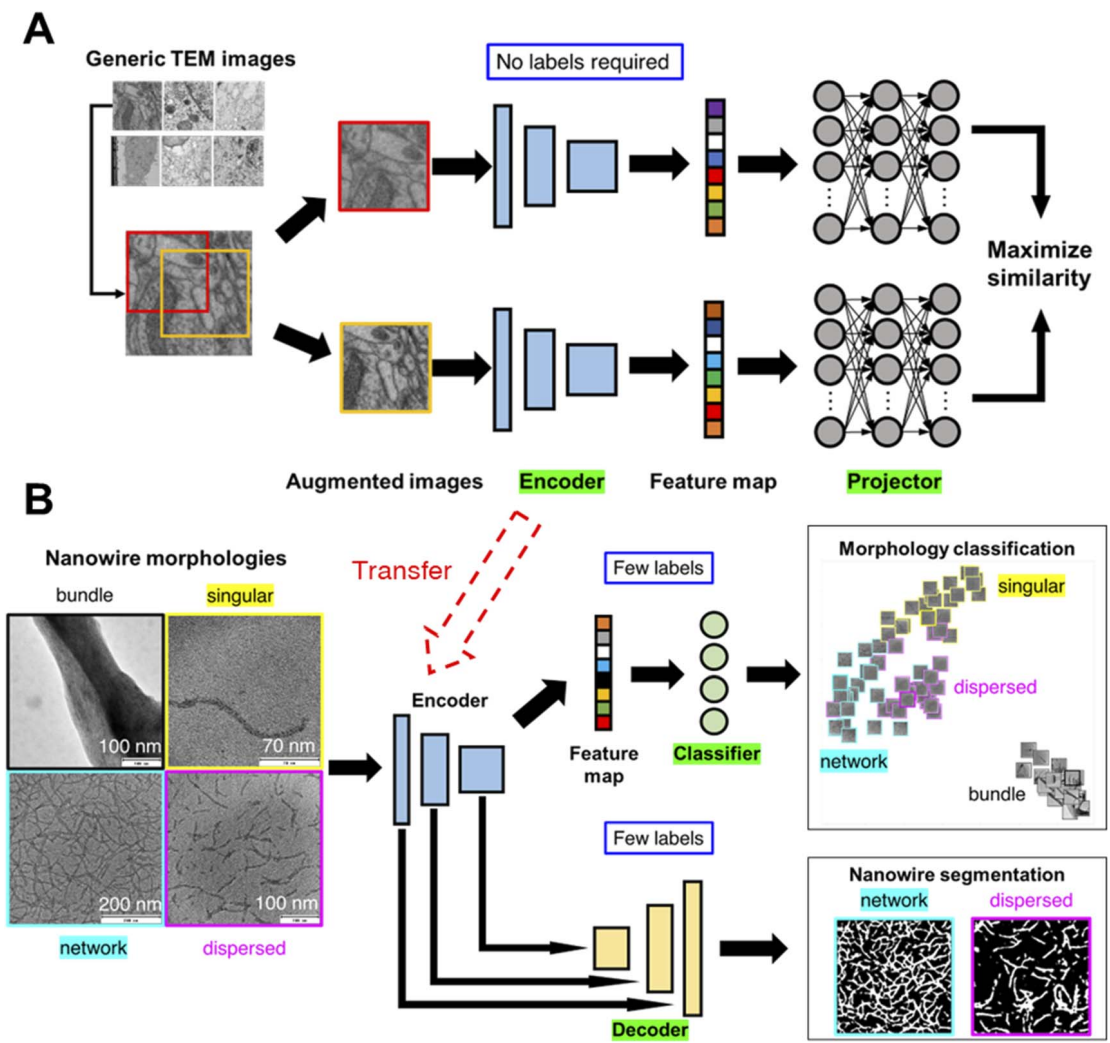

**Fig. 1** Machine learning workflow for classification and segmentation of microscopy images. Two-step generalized semi-supervised machine learning workflow for microscopy image learning. An image encoder is first trained on generic microscopy images[56] (permission from eLife Sciences Publications, Ltd, UK under CC-BY license) without supervision, *i.e.*, self-supervised. Then, the self-supervised image encoder is transferred to convert task-specific microscopy images into feature maps which are used to train multiple models for downstream tasks (*e.g.*, classification, segmentation) (A) for each image in the generic TEM image dataset, two randomly augmented images are generated. To maximize the similarity between the two augmented images, an encoder (ResNet50 (ref. 58)), followed by a projector made of three fully connected neural layers, are trained. (B) The encoder is transferred as is for the task-specific image encoding (*i.e.*, to convert TEM images of nanowire morphologies to feature maps) followed by supervised, label-efficient training of classifier and decoder for downstream tasks – classification and segmentation of TEM images.

**Nanowire morphology classification – one-shot learning**

As we observe large fluctuations in the accuracy of linear classifiers trained with only one labeled image per class (*i.e.*, one-shot learning), we want to understand how the selected labeled images or the "anchor" images affect accuracy. Using feature maps obtained from the Barlow-Twins-TEM encoder, we show one example of "good anchors" and "bad anchors" each chosen posteriorly based on the highest and lowest accuracies (Fig. 3A–B). We use t-distributed Stochastic Neighbor Embedding (t-SNE)[64] to visualize the feature maps of the test images projected in 2-dimensional space. From the t-SNE plots, we see that while there are few misclassifications between the dispersed and network morphologies when the linear classifier is trained on "good anchors" (Fig. 3C and G), most images in dispersed morphology are misclassified as network morphology when the linear classifier is trained on "bad anchors" (Fig. 3F and J). To explain the visible difference in the performance of linear classifiers trained on different "anchors", we look at the distribution of the nanowire pixel density, *i.e.*, percentage of "nanowire pixels" over all pixels, of the ground truth (*i.e.*, manually labeled images with nanowire pixels and background pixels) for test images in dispersed and network morphologies. The nanowire pixel density of the two anchor images is on the opposite sides of that of the respective median of the two morphologies for "good anchors" (Fig. 3D), but on the same sides for "bad anchors" (Fig. 3E). We also show the statistics of all 100 sets of anchors and find that the accuracy of linear classifiers trained on "opposite side anchors" is statistically higher than that trained on "same side anchors" (Fig. 3H). We conclude that the "opposite side anchors" in our study are better approximates of the medians of the test set than "same side anchors" for having smaller absolute (anchor-to-median) distance as shown in (Fig. 3I), thereby leading to better accuracies.

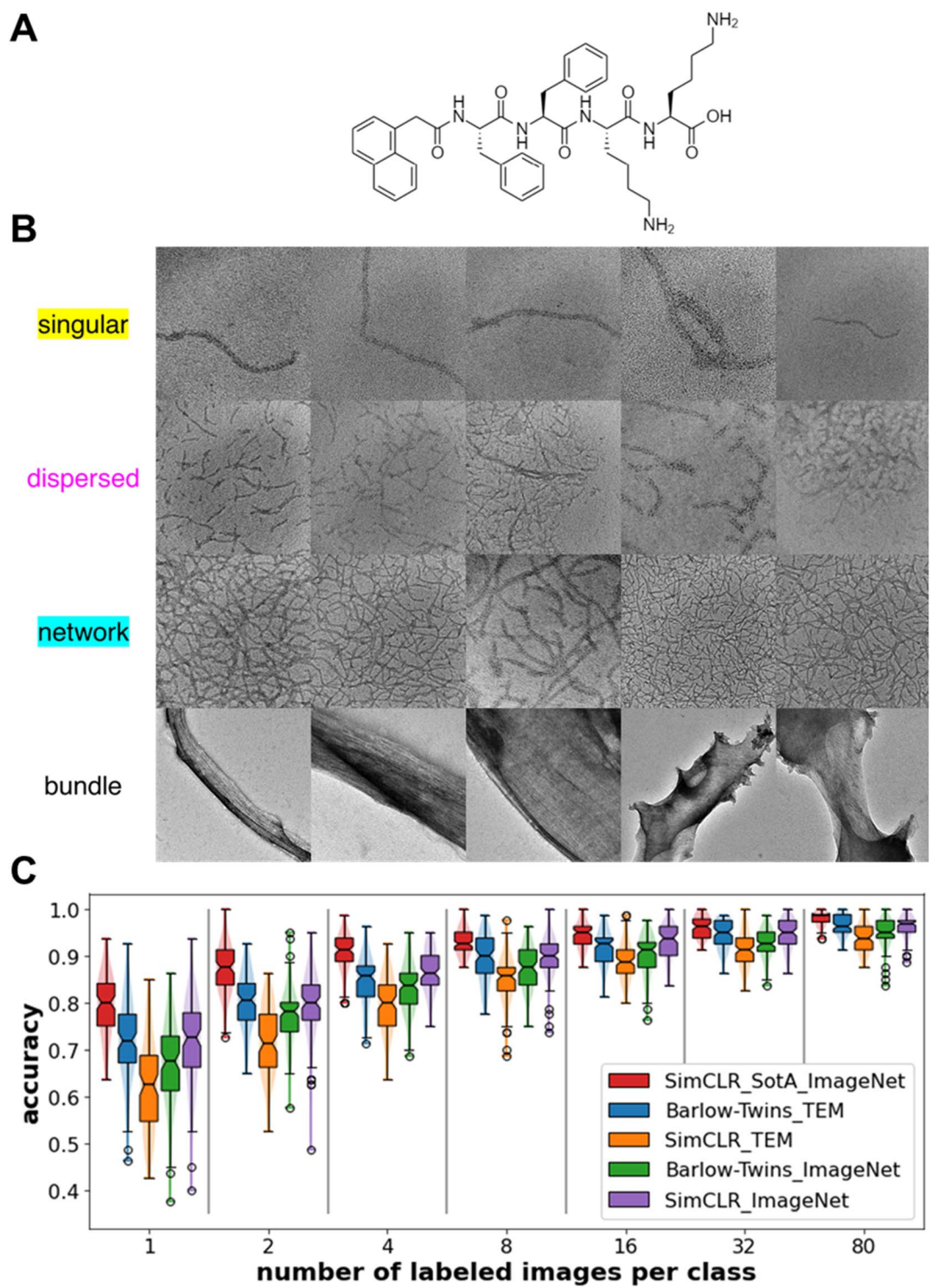

Fig. 2 Nanowire chemical composition, TEM images of nanowire morphologies, and performance of the trained model for classification task. (A) *N*-[2-(1-Naphthalenyl)acetyl]-ʟ-phenylalanyl-ʟ-phenylalanyl-ʟ-lysyl-ʟ-lysine (NapFFKK) oligopeptide structure that self-assembles into singular, dispersed, and network nanowire morphologies from water and organic solvents. (B) Representative TEM images of the four types of nanowire morphologies labeled in the figure; apart from the three oligopeptide nanowire morphologies, the bundle morphology is obtained from pilA protein nanowires (amino acid sequence of pilA: FTLIELLIVVAIIGILAAIAIPQFSAYRVKAYNSAASSDLRNLKTALESAFADDQTYPPES) harvested biologically from *Geobacter sulfurreducens* and assembled in organic solvents. (C) Classification accuracy of the trained downstream linear classifiers as a function of number of labeled images used during training of the downstream task. Legend denotes the self-supervision method and the generic image dataset used to train the encoder. Sample size is 100 for each boxplot. Notch of the boxplots indicates 95% confidence interval around the median.

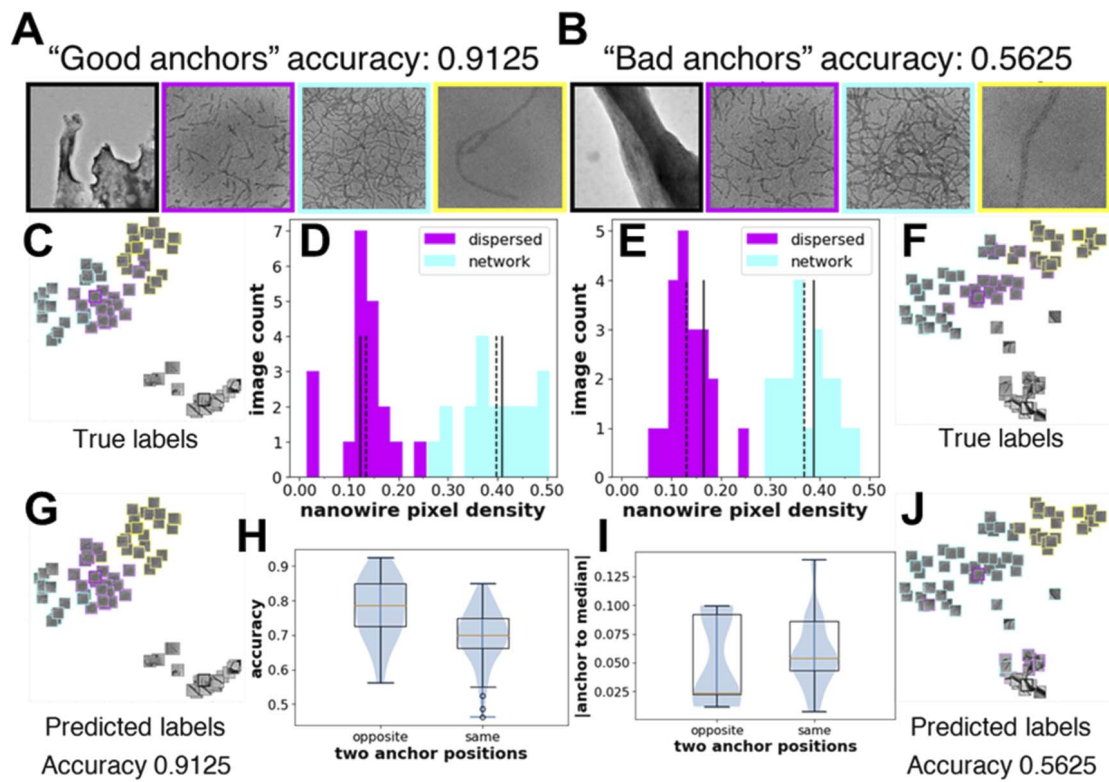

**Fig. 3** Knowledge of the underlying distribution of the images can help determine "good anchor" images for high accuracy one-shot learning. (A and B) "good anchor" images *i.e.*, a set of labeled images from each morphology, which when used for training give high classification accuracy. "Bad anchor" images which when used for training give low classification accuracy. t-SNE[64] representations of the test set colored by their true labels and by their predicted labels with "good anchor" images as training set (C and G) and "bad anchor" images as training sets (F and J). Images of the t-SNE plots of (part C, G, F, and J in original size and resolution) are provided in the ESI as Fig. S1–S4.† The image count distribution with nanowire pixel density obtained from the manual nanowire labels for the dispersed and network morphology images in the test set with "good anchors" (D) and "bad anchors" (E) images as training sets, respectively. Solid lines are positions of the two "anchors", and dashed lines are positions of the two medians. Test set size is 20 for both dispersed and network morphology. (H) The prediction accuracies with different relative positions of the two "anchors" to the median of the dispersed and network images in the test set. For the 100 samples obtained in Fig. 2C for Barlow-Twins-TEM, 25 resulted on the opposite side, 75 resulted on the same side. (I) The sum of the absolute distance between the anchor and median of the two morphologies with different relative positions of two "anchors" to the median of the dispersed and network images in the test set.

### Nanowire segmentation

Next, we tackle the task of segmentation of nanowires for the dispersed and percolated morphologies. We calculate both the Dice score (eqn (1)) and the Intersection-over-Union (IoU) score (eqn (2)) for each prediction. For assessing the performance of segmentation models, pixel-level classification is the basis for the two segmentation scores. On a pixel level, nanowire pixels predicted as nanowire pixels are regarded as true positives (TP), nanowire pixels predicted as background pixels are regarded as false positives (FP), background pixels predicted as background pixels are regarded as true negatives (TN), background pixels predicted as nanowire pixels are regarded as false negatives (FN).

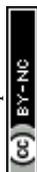

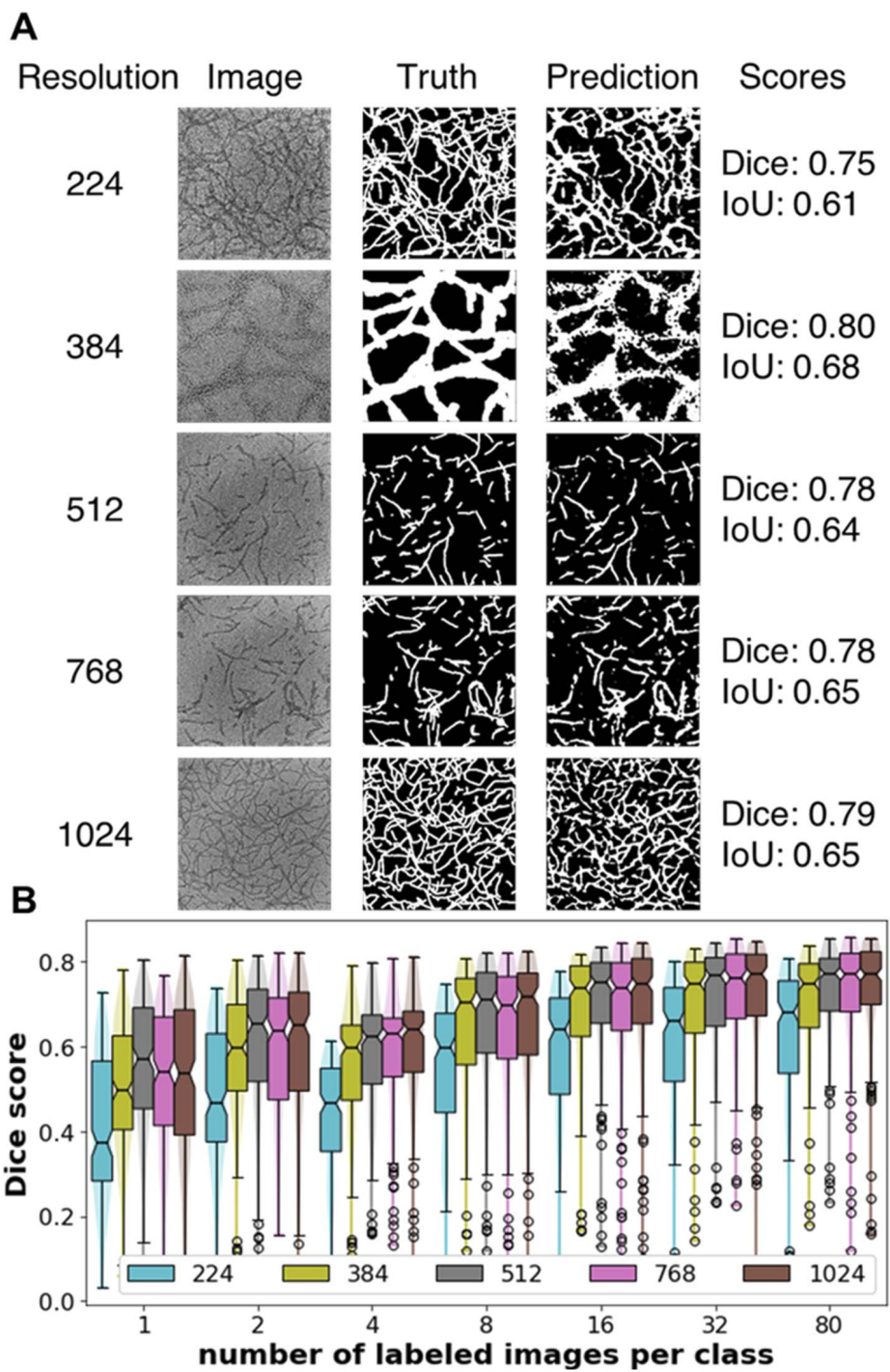

**Fig. 4** Nanowire segmentation task performance. (A) Original image, manually labeled nanowires, predicted nanowires, and segmentation performance Dice scores and Intersection-over-Union (IoU) scores at five input image resolutions. (B) Dice scores for the nanowire segmentation task. Legend indicates the resolution of the input images. Sample size is 200 for each boxplot. Notch of the boxplots indicates 95% confidence interval around the median.

The Dice score is the ratio of two times the intersection of the predicted and the true nanowire pixels over the sum of the number of the predicted and the true nanowire pixels. Dice score is calculated as:

$$\text{Dice} = \frac{2TP}{2TP + FP + FN} \qquad (1)$$

The Intersection-over-Union (IoU) score is the ratio of the intersection of predicted and true nanowire pixels over the union of the predicted and the true nanowire pixels. The IoU score is calculated as:

$$\text{IoU} = \frac{TP}{TP + FP + FN} \qquad (2)$$

For any image, the two metrics are always positively correlated, *i.e.*, if model A is better than B under one metric, it is also better than model B under the other metric.

Our U-Net model with transferred encoder (trained on resolution of $512 \times 512$ unlabeled generic TEM images with the Barlow-Twins method) works well with images of resolutions up to $1024 \times 1024$ (Fig. 4A). For segmentation, images containing nanowires present a more difficult problem compared to images with isolated small nanoparticles due to larger intersection area between the nanowires and the background. Of the five input resolutions, our U-net model trained with images of resolution 224 underperforms the higher resolution images likely due to poor contrast when the images are resized to such low resolution. We observe a plateau in the Dice score from 8 to 80 labeled images per class and a drop-off from having 8 down to 4 labeled images per class (Fig. 4B). With transferred encoder, our Unet model can achieve good performance (median Dice score > 0.70) with just 8 labeled images per class for training, less than half of the number of test images (20 per class). IoU scores follow the same qualitative trend as Dice scores; the values of IoU scores are always smaller than Dice scores (Fig. S5†). Encoder trained with unlabeled images of different resolutions give statistically similar Dice and IoU scores (Figs. S6 and S7†). With the segmentation of nanowires as an example, we show that trained encoder can capture not only global information that was important for image-level tasks such as morphology classification but also local information relevant to pixel-level tasks such as segmentation.

### Broader applicability of the machine learning workflow

Next, we test the generalizability of our machine learning workflow on classification of nanoparticle morphologies. We use TEM images in the AutoDetect-mNP dataset[65] which was originally used for shape analysis of gold metal nanoparticles. Here we use those images for a new task – to classify the morphologies that the assembled nanoparticles adopt regardless of nanoparticle shape (short or long nanorods or triangular prisms). This repurposed mNP dataset for this assembled nanoparticle morphology classification task contains three morphology categories: dispersed nanoparticles (Fig. 5A), separated clusters (Fig. 5B), and percolating cluster (Fig. 5C), each category contains 100 images. With the same training protocol applied in previous sections for peptide/protein nanowire TEM images, *i.e.*, training and predicting with linear classifiers on feature maps obtained from encoders trained with self-supervised methods, we report the classification accuracies from the linear classifier on the feature maps of the repurposed mNP dataset's test sets (Fig. 5D). We observe that the accuracies obtained from feature maps trained on both self-supervised methods are comparable to that obtained with the state-of-the-art encoder (SimCLR[47] method trained on full ImageNet[36] dataset). We note that gold nanoparticles have higher contrast with the background compared to peptide/protein nanowires with the background in TEM, therefore images in the mNP dataset are subject to less background noise. Here, we have shown that with the focus on morphology and not on the shape of the nanoparticles, self-supervised encoders trained with significantly fewer number of images can achieve comparable accuracies with the state-of-the-art encoders trained with the ImageNet dataset.

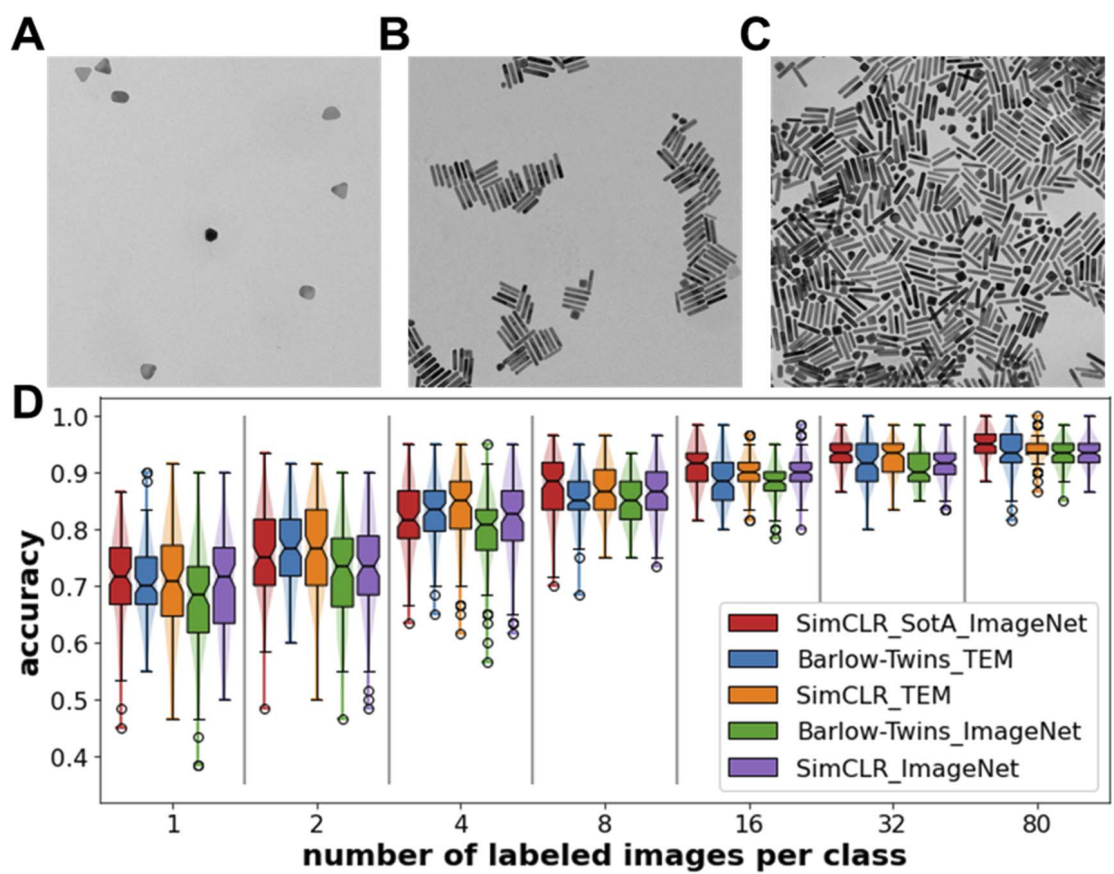

**Fig. 5** Morphology classification performance on mNP dataset. Representative TEM images of the three metal nanoparticle morphologies (A) dispersed nanoparticles, (B) separate clusters, (C) percolating cluster from the AutoDetect-mNP dataset, DOI: https://doi.org/10.6078/D1WT44 and DOI: https://doi.org/10.6078/D1S12H. (D) Classification accuracy of the trained downstream linear classifiers as a function of number of labeled images used during training of the downstream task. Legend denotes the self-supervision method and the generic image dataset used to train the encoder. Sample size is 100 for each boxplot. Notch of the boxplots indicates 95% confidence interval around the median.

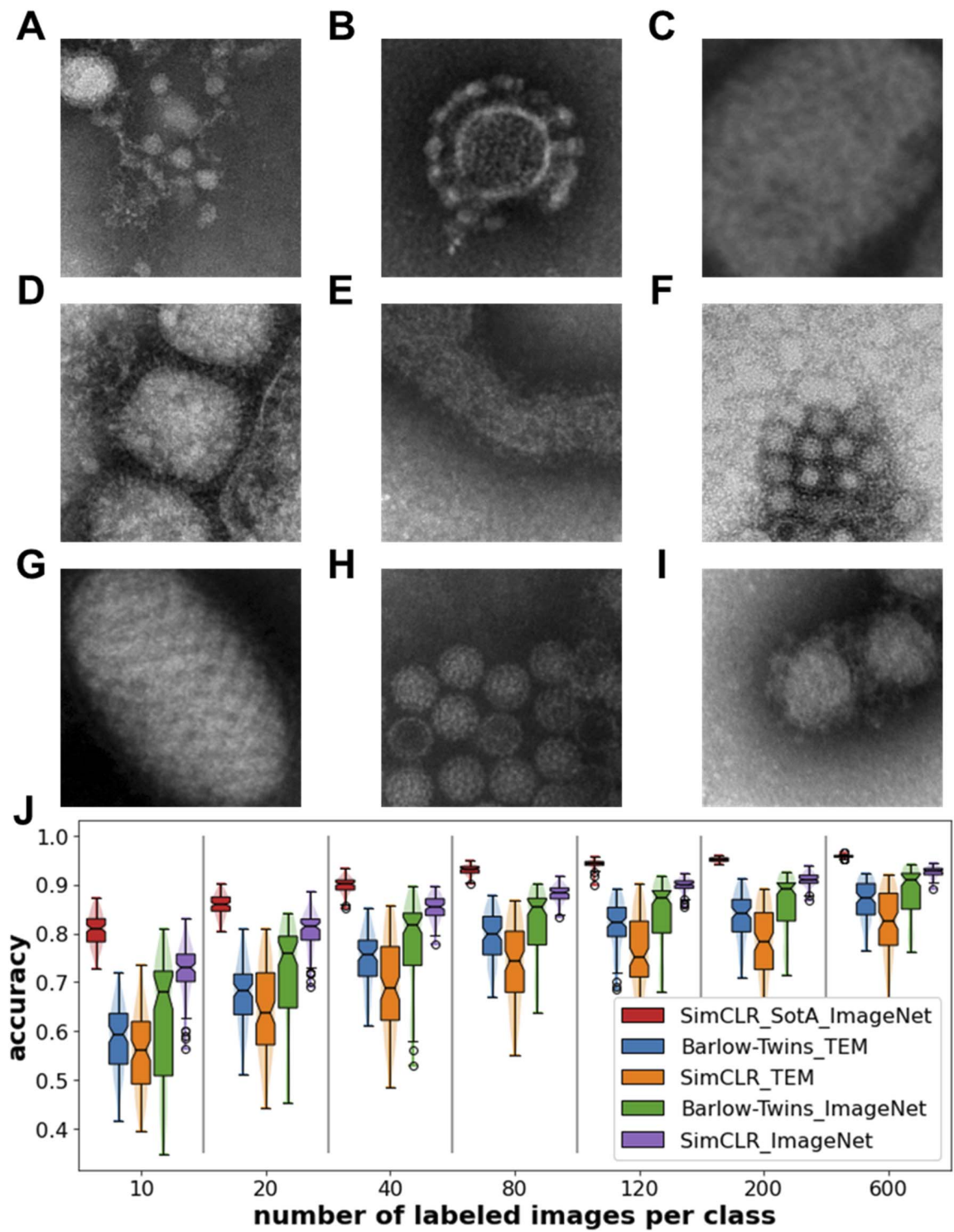

Fig. 6   Virus identification performance on TEM virus dataset. Representative TEM images of the different virus types (A) Astrovirus, (B) Nairovirus, (C) Coxpox, (D) Influenza, (E) Nipah, (F) Norovirus, (G) Orf, (H) Papilloma, (I) Rift Valley from the TEM virus dataset[66] DOI: https://doi.org/10.17632/x4dwwfwtw3.3. (J) Classification accuracy of the trained downstream linear classifiers as a function of number of labeled images used during training of the downstream task. Legend denotes the self-supervision method and the generic image dataset used to train the encoder. Sample size is 100 for each boxplot. Notch of the boxplots indicates 95% confidence interval around the median.

Next, we test our machine learning workflow for the automated task of identifying virus from TEM images. We use an open-access TEM virus dataset[66] that contains a diverse set of TEM images of more than 10 types of viruses. We choose 9 types from the dataset for the purpose of our test: Astrovirus (Fig. 6A), Nairovirus (Fig. 6B), Coxpox (Fig. 6C), Influenza (Fig. 6D), Nipah (Fig. 6E), Norovirus (Fig. 6F), Orf (Fig. 6G), Papilloma (Fig. 6H), Rift Valley (Fig. 6I). These TEM images were parts of whole-slide images of a virus or viruses taken by domain experts that were cut into smaller images. Matuszewski and Sintorn[51] state that misinterpretation can happen for two types of viruses that look similar or for images that selected unrepresentative part of the virus or selected the background. We apply our machine learning workflow to the TEM virus dataset and report the classification accuracies of the linear classifiers trained on the feature maps (Fig. 6J). We notice that the feature maps obtained from encoders trained on generic TEM images (Barlow-Twins_TEM, SimCLR_TEM) both underperform those trained on ImageNet images. A possible explanation is that encoders trained on generic TEM images are less tolerable to noisy dataset. Another explanation is that the ImageNet dataset was originally and specifically prepared for object identification with each image labeled with an object's name while the CEM500k dataset of TEM images were prepared with no labels on the images for general purpose of self-supervised learning. The virus identification task is similar to the original ImageNet object identification task therefore self-supervised models trained on ImageNet images give higher accuracies.

## Conclusions

In summary, we have developed a semi-supervised transfer learning workflow and demonstrated its efficacy when applied to the task of learning nanoscale morphologies from small microscopy image datasets and minimum number of manual labels. We show that our encoder trained with <1000 unlabeled images achieve comparable performance with state-of-the-art encoder trained with more than one million images. Our downstream task models (*e.g.*, classification of nanowire morphology and segmentation of nanowires) trained on the encoded feature maps can achieve >90% accuracy on classification of nanowire morphologies and >0.70 Dice score on segmentation of nanowires training with <10 labeled images per class. With knowledge of the underlying image distribution of the two morphologies that are harder to visually distinguish, we show that it is possible to achieve >90% accuracy training with just one labeled image per morphology.

We also show that broader applicability of our machine learning workflow for classification and identification tasks in other microscopy images (*e.g.*, assembled nanoparticles of various shapes, viruses) with limited labeled images for training. While there may exist actionable qualification criteria for manual labeling an image for an object identification task, subtle morphological differences are intrinsically harder for human experts to discern and classify into categories. Our machine learning workflow is precisely targeting such morphology classification problems to mitigate human biases. In addition, we also want to emphasize that thoughtful categorization and proper labeling of image data is crucial regardless of dataset size[10] and is especially important for data-limited image learning problems.

The machine learning microscopy image analysis workflow we have presented in this article should enable many fundamental studies in soft-, nano-, and bio- materials through democratizing (by using only a few labeled images in training) and automating the structural analysis for feature extraction, morphology classification, and segmentation tasks. For example, one potential application of our machine learning workflow could be real-time analysis of *in situ* microscopy images obtained at regular time intervals during thermal/solvent annealing of polymer nanocomposites when the nanoparticle morphology in the matrix polymers evolves. Another example could be automated analysis of microscopy images obtained during (shear, temperature, solvent) processing of block-copolymer films leading to phase transitions from one structure to another. With advances in high-throughput experimentation and characterization instrumentation, machine learning workflows such as the one in this article will be valuable in accelerating materials innovation and establishing molecular design–structure relationship.

## Methods

### Chemicals and materials

Fmoc-Lys(boc)-OH, Fmoc-Phe-OH, diisopropylethylamine (DIPEA), hexafluorophosphate benzotriazole tetramethyl uronium (HBTU), 1-hydroxybenzotriazole monohydrate (HOBt) piperidine, and 2-chlorotrityl chloride resin was purchased from Creosalus and used as received. Dichloromethane (DCM), *N*,*N*-dimethylformamide (DMF), methanol (MeOH), hexanes, and diethyl ether were purchased from Fischer Scientific; the DCM was distilled over calcium hydride prior to use. Isopropyl alcohol (IPA), anhydrous *N*,*N*-dimethylformamide, trifluoroacetic acid (TFA), and 1-naphthaleneacetic acid (NapAcOH) were purchased from Sigma Aldrich and used as received. 2,2,2-Trifluoroethanol was purchased from Oakridge Chemical and used as received. Deuterated dimethyl sulfoxide (DMSO-$d_6$) was purchased from Cambridge Isotopes.

### Instrumentation

Transmission electron microscopy (TEM) was performed on an FEI Technai T12 electron microscope using samples prepared on 400 square mesh carbon-coated copper grids (Electron Microscopy Sciences) at 120 kV accelerating voltage.

### Synthesis of N-[2-(1-Naphthalenyl)acetyl]-L-phenylalanyl-L-phenylalanyl-L-lysyl-L-lysine (NapFFKK)

NapFFKK was synthesized in an analogous manner to that used by Laverty, et al.[67] with the detailed procedure below using conventional solid-phase peptide synthesis protocols. 2-Chlorotrityl chloride supported PS resin was loaded into a peptide vessel, weighed, and swollen in DCM (distilled over $CaH_2$ prior to use). The solution was filtered, and a solution of Fmoc-Lys(boc)-OH and DIPEA in DCM while agitating with $N_{2(g)}$ for 2 h. The solution was drained, and the resin was washed three times with DCM before capping unreacted chlorotrityl chloride groups with an 80 : 15 : 5 solution of DCM : MeOH : DIPEA for 1 h. The solution was filtered, rinsed three times with DCM, three times with DMF, three times with hexanes, three times with IPA, three times with MeOH, and three times with again with DCM, before drying under vacuum overnight. The peptide vessel was weighed to check capping (complete to manufacturer's specified 1.6 mmol $g^{-1}$), swollen in DCM, rinsed three times with DMF, and deprotected using a 3 : 1 DMF : piperidine solution (5 min and 20 min). The remaining amino acids and naphthalene acetic acid were successively coupled to the functionalized resin using HOBt and HBTU as coupling agents, in the following order: Fmoc-Lys(boc)-OH, Fmoc-Phe-OH, Fmoc-Phe-OH, NapAcOH. For each coupling reaction, the amino acid or capping group (3 eq.), HBTU (3 eq.) and HOBt (3 eq.) were dissolved in anhydrous DMF, before DIPEA (7 eq.) was added to the solution, mixed, and immediately added to the peptide vessel. The Erlenmeyer flask used was rinsed with ~10 mL DMF, which was added to the peptide vessel, and the peptide solution was agitated under $N_{2(g)}$ for 2 h. The resin was then filtered, washed three times with DMF, and Fmoc groups removed as previously described. Following the final addition of NapAcOH, a solution of TFA : TIPS:$H_2O$ (96 : 2 : 2) solution was added in three parts over 3 h, filtering between each addition. The filtrate was concentrated under $N_{2(g)}$ or via rotational evaporation, precipitated in ether, and centrifuged. Extensive rinsing with ether was employed to reduce the TFA content; the product was then dried overnight under vacuum and isolated as a white powder.

### Bundled nanowire images generation

PilA-based protein nanowires were isolated from E. coli, purified, and bundled in accordance with the procedure described by Sun, et al.[68] and are described in detail below. Protein nanowires were harvested from Geobacter sulfurreducens expressed from E. coli using physical shearing followed filtration and collection in MilliQ water. An aliquot of the aqueous nanowire mixture (30 μL aliquot containing 0.535 μg protein per μL) was to a glass vial and dried under a stream of $N_{2(g)}$. Organic solvent (cyclohexane, THF, DMF, or acetone) was added to a final concentration of 0.10 μg nanowires per μL of solvent; the mixture was vortexed 5 times for ~1 second at high power then allowed to settle for ~20 minutes. Samples were vortexed once before transferring 5 μL via micropipette to an oxygen plasma-treated substrate (400 mesh, 3–4 nm carbon coated copper TEM grids) and drying in air, either to full solvent evaporation or for 5 min, and residual solvent wicked dry using filter paper (in the case of DMF). The TEM samples were stained for 20 s with 4 μL of a 2 wt% aqueous uranyl acetate stain, wicked dry using filter paper, rinsed three times using water droplet on parafilm method and wicked dry after each rinse, and imaged.

### Single nanowire images generation

NapFFKK was dissolved in water to a concentration of either 0.1 mg mL$^{-1}$ or 0.05 mg mL$^{-1}$, vortexed until fully dissolved, and allowed to stand for 1 h. The sample was then vortexed for 0.5 – 1 s twice and 4 μL was transferred to an oxygen plasma treated substrate (400 mesh, 3–4 nm carbon coated copper TEM grids) and drying in air for 5 min before wicking dry using filter paper. The TEM samples were stained for 40 s with 4 μL of a 2 wt% aqueous uranyl acetate stain, wicked dry using filter paper, rinsed three times using water droplet on parafilm method and wicked dry after each rinse, and imaged.

### Dispersed and network nanowire images generation

NapFFKK was dissolved in solvent (water, IPA, Acetone, ACN) with an initial concentration of 0.2 mg mL$^{-1}$, vortexed until fully dissolved (except in the case of IPA and ACN, which incompletely dissolved the peptides), and allowed to stand for 1 h. Water and acetone samples were then vortexed for 0.5 – 1 s twice prior to transferring to TEM grids; ACN and IPA aliquots were removed without vortexing to avoid resuspending non-dissolved peptides. In all cases, 4 μL was transferred to an oxygen plasma treated substrate (400 mesh, 3–4 nm carbon coated copper TEM grids) and drying in air for 5 min (the water sample was wicked dry using filter paper; organic solvents fully evaporated over this time). The TEM samples were stained for 40 s with 4 μL of a 2 wt% aqueous uranyl acetate stain, wicked dry using filter paper, rinsed three times using water droplet on parafilm method and wicked dry after each rinse, and imaged.

### Image data preprocessing

For the generic TEM images, we took a subset of 10 000 images from the CEM500K microscopy image dataset[56] and randomly selected 832 images from the subset each time we train on generic TEM images. The generic TEM images are of resolution 224 × 224.

For the generic everyday images, we took one image from each class of the ImageNet1k dataset[36] resulting in 1000 images of different classes as we do not want the images to be similar. We randomly selected 832 images from the subset each time we train on generic everyday images. The generic everyday images are in color and of resolution 224 × 224. We do not use the original labels of the images from the ImageNet dataset.

For the task-specific images of peptide/protein nanowire morphologies, 100 images were chosen for the dispersed, network, and bundle morphologies. Single nanowires require more dilution and are trickier to obtain. Due to difficulty in capturing single nanowires in the morphologies, we chose 25 images containing a single nanowire. We augment the singular morphology with a similar method used recently for under-represented carbon nanotube morphologies.[11] 75 images were generated by rotating the original 25 single nanowire image by 90, 180, and 270° resulting in a total of 100 images for the singular morphology. The morphology images are taken at different magnification, ranging from 21k× to 400k×. All the images of nanowire morphologies are at resolution 2048 × 2048.

Due to difficulty in distinguishing the network morphology from dispersed morphology in some cases, we manually labeled the nanowires in images with dispersed and network morphology to provide quantitative basis for the qualitative morphology class labels of the two easy-to-confuse morphology classes. We labeled the nanowires with masks colored in blue (to distinguish from the contents in the original grayscale image) through Microscopy Image Browser (MIB),[63] a MATLAB-based annotation software, and saved a binary image with the manual labels. Nanowires that we manually masked out with "colored" masks were labeled as "nanowire pixels", other pixels are labeled as "background pixels". We then performed DBSCAN,[69] a clustering algorithm implemented in scikit-learn package,[70] on the manually labeled "nanowire pixels". For each image with clusters of nanowires found by DBSCAN, we quantify percolation by checking whether there exists a cluster that spans both the horizontal and vertical dimension, i.e., two-dimensional percolation. To check criteria of spanning both dimensions, for each cluster, we check if the coordinate of the rightmost pixel minus that of the leftmost pixel is no less than the horizontal dimension minus two, same for the vertical dimension. We have confirmed that all the network images are percolated and all the dispersed images are not percolated. We acknowledge that the definition of percolation, in this case, is local to the image, and not necessarily representative of the material as a whole.

From the images with manually labeled nanowire masks of the dispersed and network morphologies, we created the segmentation ground truth binary maps in different resolutions, i.e. 224 × 224, 384 × 384, 512 × 512, 768 × 768 and 1024 × 1024. The binary maps are the standard truth and prediction target for binary segmentation task (for multi-class segmentation tasks, the binary map extend to the number of object class plus one, (for background)). In the binary map, nanowire pixels were given value of 1, whereas background pixels were given value of 0. The distribution of nanowire pixel density i.e., percentage of "nanowire pixels" over all pixels in a segmentation ground truth binary map, of both dispersed and network morphologies was obtained from binary maps of resolution 224 × 224 (Fig. S8†).

For the metal nanoparticle (mNP) dataset, we selected TEM images of nanoparticles from the mNP dataset.[65] The mNP dataset contains TEM images of short or long nanorod assemblies and triangular prism assemblies. The MIB software[63] was used to convert the images of file type .dm4 to .jpg. After inspection of the TEM images, we categorized the assembled nanoparticles into three categories: dispersed nanoparticles, separate clusters, and percolating cluster, with each morphology containing 100 images. The mNP images are of resolution 2048 × 2048.

For the TEM virus dataset,[66] we used 9 of the 14 virus datasets (Astrovirus, Nairovirus, Coxpox, Influenza, Nipah, Norovirus, Orf, Papilloma, Rift Valley). Adenovirus, Ebola, Lassa, Marburg, and Rotavirus datasets were the 5 datasets not used because these were easily confused across the categories judging by the confusion matrix provided in the original paper.[51] The TEM virus images are of resolution 256 × 256.

### Description of self-supervised training of image encoder

A self-supervised encoder training process consists of four main components: image augmentation, image encoder, projector, and the loss function. We implemented two self-supervised training methods: SimCLR[47] and Barlow-Twins.[57] The two methods only differ in the loss function. For both methods, batch size (i.e., the number of images to train at a time) is an important parameter. Larger batch sizes often lead to higher performance and shorter training time. However, batch size is usually limited by the GPU memory available. We chose batch sizes of 16, 32, and 64 to tune during hyperparameter tuning. All self-supervised training of the encoder was done on a single Nvidia P100-PCIE GPU provided by Google Colab Pro subscription.

For image augmentation, we performed random cropping, random left-right flip, and random color jittering. The combination of random cropping and random color jittering was shown in the original implementation of SimCLR trained on everyday images to give better performance compared to other augmentation combinations. However, TEM images, being in grayscale, are expected to be more impacted by random cropping. Thus, we chose the method of random cropping as a hyperparameter to tune. We first chose the side length of the cropped image to be either $\frac{1}{4}$ or $\frac{1}{2}$ or a random percentage between $\frac{1}{4}$ and $\frac{1}{2}$ of the side length of the original image. Then the left-bottom point of the cropped image was randomly picked so that the cropped image is not out-of-bounds of the original image. Finally, the cropped image was resized to the resolution of the original image. Random left-right flip is set to a probability of 0.5. Random color jittering consists of random hue, random contrast, and random saturation. The hue, contrast, and saturation of the image were shifted (each separately with probability of 0.8 and in random order) by a random number applied on all the pixels. For hyperparameter tuning, we also tested a model where we do not crop the image and only perform random left-right flip and random color jittering.

The encoder took in the augmented images and distilled the information into feature maps. We used ResNet50 as the encoder architecture. The ResNet50 was initialized with weights trained on ImageNet images as initializing the encoder with transferred weights gives higher performance.[71] We removed

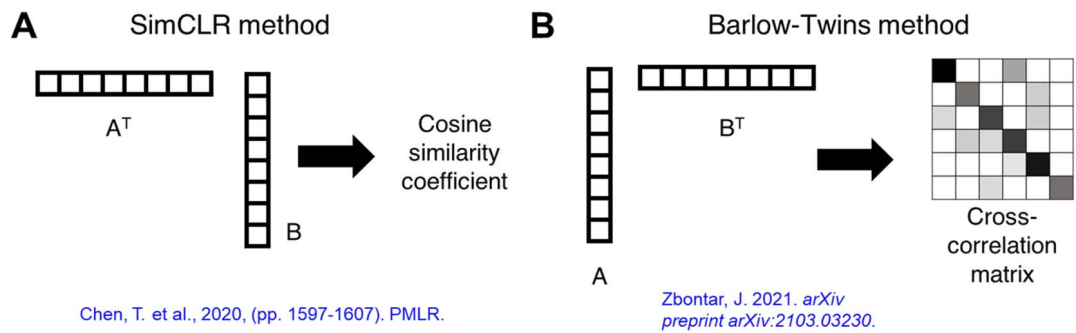

**A** SimCLR method

$A^T$

B

Cosine similarity coefficient

Chen, T. et al., 2020, (pp. 1597-1607). PMLR.

**B** Barlow-Twins method

$B^T$

A

Cross-correlation matrix

Zbontar, J. 2021. arXiv preprint arXiv:2103.03230.

**Fig. 7** Differences in vector multiplication when calculating the loss functions for SimCLR and Barlow-Twins methods. (A) With SimCLR method, a cosine similarity coefficient is obtained from pairs of projections. (B) With Barlow-Twins method, a cross-correlation matrix is obtained from pairs of projections.

the classification layer and keep the weights of the encoder to be trainable during training. The output of the encoder, *i.e.*, feature map was of dimension 2048.

The projector was constructed with three linear layers with number of nodes: 128, 64, 1024. The number of nodes in the final layer was also considered as a hyperparameter. All three layers were followed by a batch normalization layer. The first two layers also had rectified linear units as activation layer. The output of the projector is called the projection.

For each batch of images in training, two augmented images were generated from each of the original images. The projections of the images were used as inputs in the loss function. Two projections undergo row vector × column vector multiplication and obtain a cosine similarity coefficient (SimCLR method) or undergo column vector × row vector multiplication to obtain a cross-correlation matrix (Barlow-Twins method), as illustrated in Fig. 7. The loss function for SimCLR method consists of two terms: a similarity term measuring the L2-normalized cosine similarity coefficient of a "true" pair of projections (coming from the same original image), and a contrast term measuring the L2-normalized cosine similarity coefficient of a "false" pair of projections (coming from different images) as shown in eqn (3). The loss function for Barlow-Twins method also consists of two terms: an invariance term in the form of L2-normalized sum-of-squares penalizing the diagonal values in the cross-correlation matrix for deviating from unity, and a redundancy reduction term measuring the L2-normalized sum-of-squares of off-diagonal values in the cross-correlation matrix as shown in eqn (4).

$$\mathscr{L}_{\text{SimCLR}} = -\sum_b \frac{\langle z_b^A, z_b^B \rangle}{\tau \|z_b^A\|_2 \|z_b^B\|_2} + \sum_b \log \left( \sum_{b' \neq b} \exp \left( \frac{\langle z_b^A, z_{b'}^B \rangle}{\tau \|z_b^A\|_2 \|z_{b'}^B\|_2} \right) \right)$$

(3)

$$\mathscr{L}_{\text{Barlow-Twins}} = \sum_b \left( \sum_i \left( 1 - \frac{\langle z_{b,i}^A, z_{b,i}^B \rangle}{\|z_{b,i}^A\|_2 \|z_{b,i}^B\|_2} \right)^2 \right.$$
$$\left. + \lambda \sum_i \sum_{j \neq i} \left( \frac{\langle z_{b,i}^A, z_{b,j}^B \rangle}{\|z_{b,i}^A\|_2 \|z_{b,j}^B\|_2} \right)^2 \right)$$

(4)

where $z^A$ and $z^B$ are the projections, $b$ indexes the sample in a batch, $i$ indexes the vector component of the projection, $\tau$ is the temperature parameter analogous to statistical mechanics, we use the recommended value of 0.10 in our trainings, $\lambda$ is the weighting factor for the redundancy reduction term, we use the recommended value of 0.005 in our trainings.

For training, we used an initial learning rate of 0.2 and the learning rate decay schedule described in the original SimCLR or Barlow-Twins paper's implementation. All trainings were done with SGD optimizer for 200 epochs.

### Protocol for hyperparameter tuning model selection of self-supervised encoders

All self-supervised training of the encoder was done on a single Nvidia P100-PCIE GPU provided by Google Colab Pro subscription. All training of classification models was done on a single Nvidia K80 GPU provided by Google Colab free version. Three important hyperparameters (batch size, crop method and final projection layer size) were tuned for optimization of the self-supervised encoders with cross-validation for model selection on the morphology classification task. Five self-supervised encoders were trained with either Barlow-Twins or SimCLR method for each of the different hyperparameter sets we explored, the difference being the random seed used to select the 832 generic TEM images of resolution 224 × 224 to train. For the classification task, we used the trained encoders to transform the morphology images into feature maps. We used five different random seeds to split the nanowire morphology images into 80% training data, and 20% test data (not used) while keeping the class distribution balanced in both training and test sets. We performed 4-fold cross-validation on the training data, *i.e.*, 60% of all morphology images as training data and 20% of all morphology images as validation data. We trained a linear classifier consisting 4 neurons to classify the feature maps of morphology images (obtained from trained encoder) into their respective morphologies. Thus, for each model hyperparameter set, we had 5 encoders × 5 data split × 4 folds = 100 linear classifiers trained giving 100 accuracies as statistics. To mitigate overfitting, we used an early stopping criterion that stops the training of the linear classifier when the validation accuracy stops increasing in 10 consecutive epochs.

Adam optimizer with default learning rate of 0.001 was used for training the linear classifier.

For batch size, we looked at batch sizes of 16, 32 and 64, with 64 being the largest batch size we can use due to limitation of the GPU memory. For the crop method, we looked at Randcrop, *i.e.*, the cropped image length is random between $\frac{1}{4}$ and $\frac{1}{2}$ of the original image length; crop25, *i.e.*, the cropped image length is $\frac{1}{4}$ of the original image length; crop50, *i.e.*, the cropped image length is $\frac{1}{2}$ of the original image length and no crop, where the generic images were not cropped during augmentation., We used 64, 256 and 1024 neurons as the size of the last projection layer.

When tuning batch size, we used the Randcrop method and a last projection layer size of 1024. For the batch size, we observed that the classification performance of feature maps obtained with encoders trained with SimCLR method had a strong dependence on batch size while that trained with Barlow-Twins method was insensitive to batch size (Fig. 8A).

When tuning crop method, we used a batch size of 64 and a last projection layer size of 1024. For the crop method, we observed that the classification performance of feature maps obtained with encoders trained with SimCLR method became worse when the crop method was more rigorous, *i.e.*, cropped image is more different from the original image, while those trained with Barlow-Twins were insensitive to the crop method (Fig. 8B).

When tuning the size of the final projection layer, we used a batch size of 64 and the Randcrop method. Both methods are insensitive to the size of the final projection layer (Fig. 8C).

Based on observations of the hyperparameter tuning process, we have chosen batch size of 64, crop method of Randcrop and last projection layer size of 1024 for Barlow-Twins method training and batch size of 64, not cropping and last projection layer size of 1024 for SimCLR method training.

For the Barlow-Twins method, we also looked at how the resolution of the images used to train the encoder impacted the classification performance of feature maps (Fig. 9). Given the insensitivity of Barlow-Twins method to batch size, higher resolution images can be used as input when training the encoder using smaller batch size without losing accuracy.

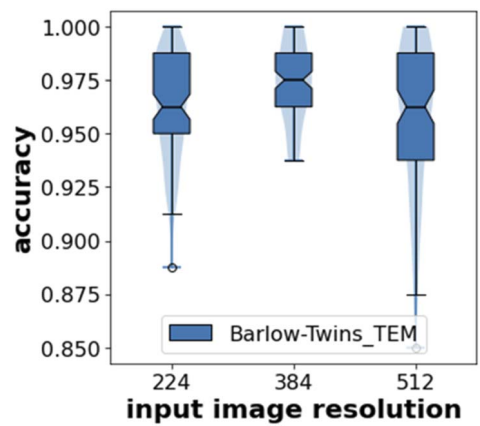

**Fig. 9** Impact of input image resolution on self-supervised training of Barlow-Twins method. Accuracies for classification model performance with feature maps obtained from the trained encoders. The batch size when training images of the three resolutions 224, 384, and 512 was 64, 16, and 8, respectively.

### Protocol for classification task performance assessment

All training of classification models was done on a single Nvidia K80 GPU provided by Google Colab free version. Except for the SimCLR_SotA_ImageNet encoder which was obtained as published from the official SimCLR repository on Github, the other four encoders were the best performing encoders selected from the hyperparameter tuning protocol.

For the classification task, we used the trained encoders to transform the morphology images into feature maps. For the peptide/protein nanowire dataset and the mNP dataset, we used different random seeds to split the nanowire morphology images into 80% training data, and 20% test data while keeping the class distribution balanced in both training and test sets. From the 80 training images per class, we then randomly selected 1, 2, 4, 8, 16, 32 or 80 images per class (with a fixed random seed of 42) as the labeled training image set. For the TEM virus dataset, we used the training, validation and test dataset provided by the original paper. We then randomly selected 10, 20, 40, 80, 120, 200 or 600 images per class (with different random seeds) as the labeled training image set. To keep the sample size the same for performance assessment, we used 100 random seeds for SimCLR_SotA_ImageNet encoder and 20 random seeds for the other four locally trained encoders because each locally trained encoders already had 5 replicate models during self-supervised training. The training procedure of the linear classifier was the same as described in the hyper-parameter tuning protocol.

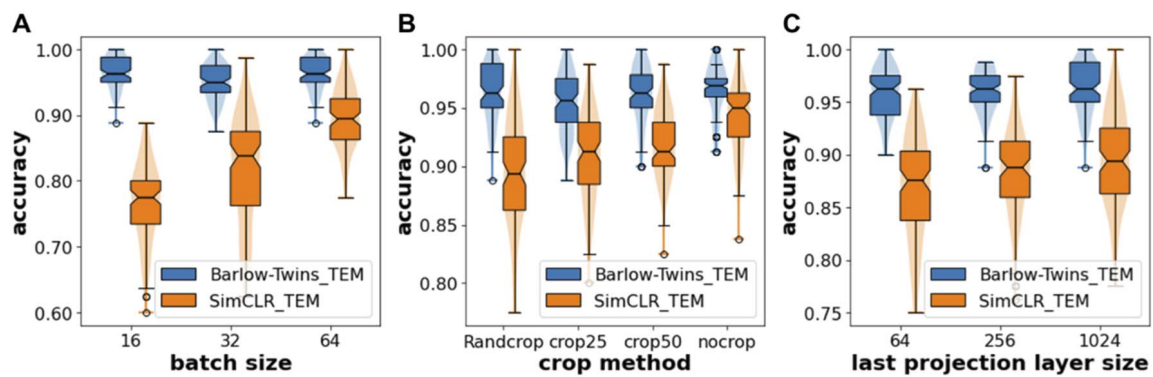

**Fig. 8** Hyperparameter tuning accuracies for classification model performance with feature maps obtained from encoders, trained with Barlow-Twins and SimCLR method, respectively, on generic TEM images. (A) Impact of batch size. Barlow-Twins method is insensitive to batch size used in training. The performance of SimCLR method, however, suffers when batch size is small. (B) Impact of crop method. Barlow-Twins method is insensitive to the crop method, but SimCLR performs better without cropping. (C) Impact of the last projection layer size. Both methods are insensitive to the size of the last projection layer.

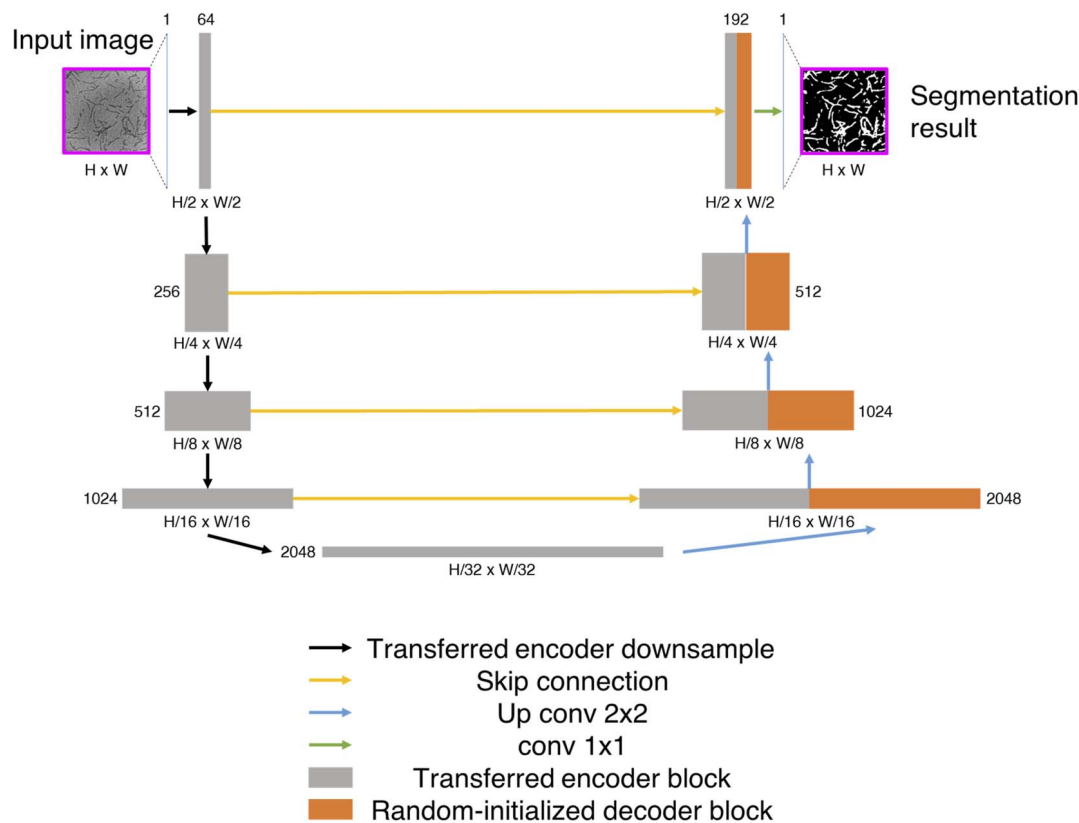

**Fig. 10** Detailed schematic of the Unet architecture with transferred encoder.

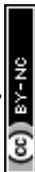

### Segmentation model description

We used the Unet architecture[59] as the segmentation model with our trained encoder as the down sampling module. Skip connections were established between the transferred encoder and a decoder with random initialized weights. The exact encoder-decoder network is shown in Fig. 10. The weights of the transferred encoder were not updated, *i.e.*, fixed as constants during training.

### Protocol for segmentation task performance assessment

All training of segmentation models was done on a single Nvidia K80 GPU provided by Google Colab free version. Compared to the classification task, the segmentation task was much slower

(2 orders of magnitude in terms of training and prediction time) due to the large overhead of the encoder-decoder model. Five encoders were trained with Barlow-Twins method on generic TEM images of resolution of either 224 × 224, 384 × 384 or 512 × 512, *i.e.*, five encoders for each resolution. For the nanowire morphology images of dispersed and network morphologies, we resized the images to resolutions of 224 × 224, 384 × 384, 512 × 512, 768 × 768 and 1024 × 1024. For the segmentation task, we used the images as inputs to the encoder-decoder network. We used one fixed random seed of 42 to split the nanowire morphology images into 80% training data, and 20% test data, while keeping the class distribution balanced in both training and test sets. We then from the 80 training images per class randomly selected 1, 2, 4, 8, 16, 32 and 80 images per class (with a fixed random seed of 42) as the labeled training image set. We used (1 − Dice score) as the loss function in training of the segmentation model. For the Dice scores or IoU scores, the sample size was 5 encoders × 40 images in the test set = 200. Adam optimizer with default learning rate of 0.001 was used for training the segmentation model. The number of epochs for models with different number of labeled training images were 120, 120, 120, 60, 40, 30 and 20 respectively.

## Data availability

The python code for implementing the machine learning models with Keras and Tensorflow is available at **https://github.com/arthijayaraman-lab/semi-supervised_learning_**

microscopy_images. The image dataset of nanowire morphologies is deposited on the open-access data repository Zenodo with DOI: https://doi.org/10.5281/zenodo.6377140. All data and models generated during and/or analyzed during the current study are available from the corresponding author upon reasonable request.

## Author contributions

S. L. devised the idea and led the machine learning workflow development with guidance and feedback from A. J.; B. M. performed the synthesis of the synthetic peptide nanowires and microscopy imaging for the synthetic peptide nanowires and bio-derived protein nanowires with feedback from T. E.; B. M. and S. L. decided on the qualification criteria of different morphologies; S. L. created the segmentation ground truth labels and the morphology class labels; S. L. wrote the python code for implementing the machine learning workflow, trained and tested the model performances; S. L., B. M., T. E., and A. J. wrote the manuscript.

## Conflicts of interest

The authors declare no competing financial interest.

## Acknowledgements

The authors acknowledge financial support from the U.S. National Science Foundation, Grant NSF DMREF #1921839 and #1921871.